\documentstyle[aps,twocolumn]{revtex}
 
\begin{document}


\newcommand{\beq}{\begin{equation}}
\newcommand{\eeq}{\end{equation}}
\newcommand{\beqa}{\begin{eqnarray}}
\newcommand{\eeqa}{\end{eqnarray}}
\newcommand{\OZ}{\overline{Z}}
\newcommand{\oz}{\overline{z}}
\newcommand{\ida}{I(X;Y\hspace{-1.3mm}\downarrow\hspace{-0.8mm} Z)}
\newcommand{\pe}{\hspace*{\fill} $\Box$\\ \ \\}

\newtheorem{theo}{Theorem}

\def\u{\uparrow}
\def\d{\downarrow}
\def\F{{\cal F}}
\def\D{{\cal D}}
\def\I{{\cal I}}
\def\half{\frac{1}{2}}
\def\ket#1{|\,#1\,\rangle}
\def\bra#1{\langle\, #1\,|}
\def\braket#1#2{\langle\, #1\,|\,#2\,\rangle}

\title{Quantum cryptography on noisy channels: quantum versus classical 
key-agreement protocols}

\author{N. Gisin$^{(1)}$ and S. Wolf$^{(2)}$}
\address{(1) Group of Applied Physics, University of Geneva, 1211 Geneva,
Switzerland}
\address{(2) Dept. of Computer Science, Swiss Federal Institute of Technology
(ETH Z\"urich),
ETH Zentrum, 
8092 Zurich, Switzerland}

\date{\today}


\maketitle

\begin{abstract}
When the 4-state or the 6-state protocol of quantum cryptography is carried
out on a
noisy (i.e.\ realistic) quantum channel, then the raw key has to be processed
to reduce the
information of an adversary Eve down to an arbitrarily low value, providing
Alice and Bob
with a secret key. In principle, quantum 
algorithms as well as classical algorithms can be used for this processing.
A natural question
is: up to which error rate on the raw key is a secret-key agreement at all
possible? Under
the assumption of incoherent eavesdropping, we find that the quantum and
classical 
limits are precisely the same:
as long as Alice and Bob share some entanglement both quantum and classical
protocols provide secret keys.
\end{abstract}



Quantum cryptography lies at the intersection of two of the major sciences
of the
20th century: quantum mechanics and information theory. Moreover, due to the
intimate
relation between quantum cryptography and quantum non-locality, another
major scientific
achievement of this century, relativity, is not far away. This letter
concerns the dialog
between quantum physics and information theory in the context of optimal
eavesdropping on
a quantum channel and the corresponding secret-key agreement that,
in principle, both quantum and classical algorithms provide.
This analysis of the interplay of these complementary approaches reveals
surprising
connections.

To provide a secure communication, quantum cryptography \cite{BB84} exploits
quantum
correlations to establish secure keys (which are then the basis for standard
information-based
cryptosystems). Alice prepares a pair of qubits (i.e. a pair of spin
$\half$) in a maximally
entangled state, sends one qubit to
Bob and keeps the other one. Then they both measure their qubit in a basis
chosen independently
at random within a set of 2 or 3 bases for the 4- and 6-state protocols,
respectively. 
(Alice could also prepare a qubit in a state compatible with one of the
bases and send it to
Bob, but the protocol with qubit pairs is equivalent and better suited for
the purpose of this
letter).
On a perfect channel the correlations are maximal and the protocol is
straightforward,
the secure key results essentially without any information theoretical
algorithm. However,
in practice the quantum channel is noisy and elaborated protocols are needed.
These require as input an upper bound on the
information accessible to the eavesdropper, a bound set by the laws of
quantum physics.
Natural questions are: up to which error rate is secure key agreement 
at all possible? This letter presents the answer to this question under 
the assumption of incoherent eavesdropping.

Below, we first review general incoherent eavesdropping
and  summarize recent results on secret-key agreement by public discussion. 
Then, the cases that Bob
has more or less information than Eve are treated.

By {\it general incoherent eavesdropping} we mean the following.
First, Eve lets each qubit sent by Alice to Bob interact with independent
ancillas. The dimension of the
ancillas and the interaction are arbitrary, except that the interaction is
described by a
unitary operator $U$:
\beqa
U\ket{\u}\otimes\ket{0} &=& \sqrt{\F}\ket{\u}\otimes\psi_\u +
\sqrt{\D}\ket{\d}\otimes\phi_\u
\label{Uu} \\
U\ket{\d}\otimes\ket{0} &=& \sqrt{\F}\ket{\d}\otimes\psi_\d +
\sqrt{\D}\ket{\u}\otimes\phi_\d
\label{Ud}
\eeqa
where the $\psi_j$ and $\phi_j$ are the normalized states of Eve's ancilla
when Bob
receives the qubit undisturbed and disturbed, respectively. The former case
happens with
probability $\F$, called the fidelity, and the latter with probability $\D$,
called the
disturbance or, equivalently, the Quantum Bit Error Rate (QBER).
Next, Eve stores her ancillas until she learns the bases used by Alice to
encode the qubits. Finally, she measures her ancillas one after the other,
using any measurement
scheme compatible with the laws of quantum physics. Clearly, if Eve
interacts only
weakly with the qubits, then she disturbs the channel only weakly, hence the
QBER is low,
but Eve gets little information. On the contrary, if the interaction is
strong, Eve gains
more information, but the QBER is larger. Optimal here means that for any
given QBER, Eve
chooses the qubit-ancilla interaction and her measurement to maximize her
information.
The QBER can be well estimated by Alice and Bob
by comparing a fraction of their (classical) bits. From this and from the
laws of
quantum physics they can bound Eve's information. 

The state vectors $\psi_j$ and $\phi_j$ in (\ref{Uu}, \ref{Ud})
have to be such that they define a unitary operator $U$. 
Moreover, we choose them such that $U$ has the same effect on all qubits
sent by Alice (all the Poincar\'e vectors are shrunk by the same factor). 
This is called symmetric eavesdropping. It simplifies the
analysis considerably. It is important to note that one can assume symmetric
eavesdropping 
without loss of generality, as proven in \cite{CiracGisin97,FGGNP}. 
Indeed, Eve can make her strategy look
symmetric to Bob by applying arbitrary rotations $R$ to the qubit
immediately before and
$R^{-1}$ immediately after it
interacts with her ancilla. As she knows which rotation she applies, she
does not lose
information. From Bob's point of view, however, the arbitrary rotations make
the disturbance
appear symmetric. From this symmetry condition one obtains the following
relation 
for the fidelity $\F$ \cite{CiracGisin97,FGGNP,Helle6state}:
\beq
\F=\frac{1+\langle \phi_\u|\phi_\d\rangle }{2-\langle \psi_\u|\psi_\d\rangle +\langle\phi_\u|\phi_\d\rangle }
\label{FF}
\eeq

The explicit form of the $\psi_j$ and $\phi_j$ and Eve's optimal
measurements are given
in \cite{CiracGisin97,FGGNP} for the 4-state protocol (BB84) and in 
\cite{Helle6state,Dagmar6state}
for the 6-state protocol (note that for the latter $\langle \phi_\u|\phi_\d\rangle =0$). 
This provides the joint probability distribution $P_{XYZ}$ of
the random variables $X$, $Y$, and $Z$
to which Alice, Bob, and Eve have access, respectively.
It turns out that Eve's random variable is composed of 2 bits
 $Z=[Z_1,Z_2]$, where $Z_1=X\oplus Y$ 
($\oplus$=xor), i.e., $Z_1$ tells Eve whether Bob received the qubit
disturbed ($Z_1=1$) 
or not ($Z_1=0$) (this is a consequence of the fact that the $\psi$ and
$\phi$ states
in (\ref{Uu},\ref{Ud}) generate orthogonal sub-spaces). 
The probability that Eve's second bit indicates the correct value
of Bob's bit depends on whether the qubit was disturbed or not:
Prob$(Z_2=Y|X=Y)=\delta_0$ and
Prob$(Z_2=Y|X\ne Y)=\delta_1$ (note that for the 6-state protocol
$\delta_1=1$). 
The relevant Shannon informations are then:
\beqa
\I_{Bob}&=&1+\F\log_2(\F)+(1-\F)\log_2(1-\F) \\
\I_{Eve}&=&\F\left(1+\delta_0\log_2(\delta_0)+(1-\delta_0)\log_2(1-\delta_0)\right
) \\ \nonumber
&+&(1-\F)\left(1+\delta_1\log_2(\delta_1)+(1-\delta_1)\log_2(1-\delta_1)\right)
\eeqa
For the 4-state protocol Eve's Shannon information is maximal when
$\delta_0=\delta_1=\half+\sqrt{\F(1-\F)}$. 
Let us concentrate on the point where $\I_{Bob}=\I_{Eve}$, at 
$QBER_0=1-\F_0=\frac{1-1/\sqrt{2}}{2}$. For QBERs below this threshold, Bob
has more 
information than Eve, while above QBER$_0$ Bob has less information than Eve. 
We shall see that in the latter case Alice and Bob can still exploit their 
authenticated classical communication channel to overcome their initial
drawback.
Note that the noise corresponding to QBER$_0$ in the 4-state protocol is
precisely
the limit above which Bell inequality \cite{CHSH} can no longer be violated 
\cite{GisinHuttner97,CiracGisin97,FGGNP}, while for the 6-state protocol it
corresponds
to the optimal Universal Quantum Cloning Machine \cite{Buzek}.

The situation that results when  the 4-state or the 6-state protocol is used,
and if all the parties obtain classical random variables by 
carrying out measurements
after each bit sent, is a special case of the more
{\it general scenario of secret-key agreement by public discussion} from common
information described by Maurer~\cite{ka}. In this setting, two parties 
Alice and Bob who are willing to generate a secret key have access to repeated
independent realizations of (classical) random variables $X$ and $Y$, 
respectively, whereas an adversary Eve learns the outcomes of a
random variable $Z$. Let $P_{XYZ}$ be the joint distribution of the three
random variables. Additionally, Alice and Bob are allowed to communicate
over a noiseless and  authenticated, but otherwise completely insecure channel.
In this situation, the secret-key rate $S(X;Y||Z)$ has been defined as the 
maximal rate at which Alice and Bob can generate a secret key that is equal 
for Alice and Bob with overwhelming probability and about which Eve has  
only a negligible amount of information (in terms of Shannon entropy).
For a detailed description of the general scenario and the secret-key
rate as well as for various bounds on $S(X;Y||Z)$,
see~\cite{ka,ittrans,techrep}.

A first result analyses the case when Bob's random variable
$Y$ provides more (Shannon-) information about Alice's $X$ than Eve's $Z$
does (or vice versa):
then this advantage can be exploited to generate a secret key:
\beqa\label{lbo}
S(X;&Y&||Z)\geq \\
&& \max\, \{I(X;Y)-I(X;Z)\, ,\, I(Y;X)-I(Y;Z)\} \nonumber
\eeqa
This bound follows from an earlier result by Csisz\'ar and K\"orner~\cite{ck}.
It guarantees the possibility of secret-key agreement using error correction and
(classical) privacy amplification whenever Bob has more information
on Alice's bits than Eve. It is somewhat surprising 
that this bound is not tight, in particular 
that secret-key agreement is even be possible 
when the right-hand side of (\ref{lbo}) vanishes or is negative. However,
it was shown that the positivity of the expression on the right-hand side
of (\ref{lbo}) is a {\em necessary\/} condition for the possibility of 
secret-key agreement by {\em one-way communication}: whenever Alice and 
Bob start in a disadvantageous situation compared to Eve, then {\em feedback\/} 
is necessary. The corresponding initial phase of the key-agreement protocol is
then called {\em advantage distillation}.

Let us come back to quantum cryptography  and first briefly apply the general
scenario
for secret-key agreement to the case when {\it Bob has more information than
Eve}.
This case is relatively simple since, before starting the classical phase of
the protocol,
Bob has already an advantage over Eve. 
Hence inequality (\ref{lbo}) guarantees that a positive secret-key rate can
be achieved
using only error correction and privacy amplification. Actually, inequality
(\ref{lbo})
provides even a lower bound on the achievable secret-bit rate.

Next, we consider the case when {\it Bob has less information than Eve}. 
This case may seem hopeless. However, there are
still two reasons to keep optimism. First, we shall see that for QBERs not
too large 
Alice's and Bob's qubits are still entangled, hence the technique of Quantum
Privacy Amplification
(QPA) \cite{QPA} can be used. Next, Alice and Bob can take advantage of the 
authenticity of the public channel
and carry out an {\it advantage distillation} protocol. We shall
successively analyze these two possibilities and show that despite their
differences, the former processing quantum information while the second is
entirely classical, both can be applied up to precisely the same maximum
QBER$_{\max}$!

The qubit sent by Alice to Bob is initially maximally entangled with another
qubit that stays in Alice's hands. Once Bob received his qubit, Alice and
Bob qubits are
in a mixed state $\rho_{AB}$. If $\rho_{AB}$ is separable, then the correlation
between Alice and Bob could be established by purely local operations and
classical (public)
discussions, hence there is no way for Alice and Bob to base a secret key on
this correlation.
If, however, $\rho_{AB}$ is entangled, then it contains some quantum (i.e.\
non-classical)
correlations. If Alice and Bob have many pairs of qubits, each in an
entangled state
$\rho_{AB}$, they can apply {\it Quantum Privacy Amplification} (also called
entanglement
distillation or purification) \cite{QPA}: after some local action involving
pairs of qubit
pairs and some public discussion,
some qubits pairs will be more entangled at the cost that
some other qubit pairs are destroyed. Repeated use of this protocol provides
Alice and
Bob with qubit pairs arbitrarily close to maximal entanglement,
as would have been obtained with a perfect channel.
They can thus be used for secret-key agreement.

According to the general incoherent eavesdropping strategy
(\ref{Uu},\ref{Ud}) and
assuming Alice prepares the qubits in the singlet state,
$\rho_{AB}$ takes the form (in the computational basis
$\{\ket{00},\ket{01},\ket{10},\ket{11}\}$):
\beq
\rho_{AB}=\half\pmatrix{\D & 0 & 0 & -\D c_\phi\cr 0 & \F & -\F c_\psi & 0\cr 
0 & -\F c_\psi & \F & 0\cr -\D c_\phi & 0 & 0 & \D} 
\eeq
where $c_\psi=\langle \psi_\u|\psi_\d\rangle $ and $c_\phi=\langle \phi_\u|\phi_\d\rangle $. Using the
Peres-Horodecki
\cite{PeresHorodecki} separability condition, one finds that $\rho_{AB}$ is
entangled if and only if
$\D>\F c_\psi$ and $\F>\D c_\phi$. Using the explicit relations among $\F$,
$\D$, $c_\psi$
and $c_\phi$ one obtains the maximum QBER$_{\max}$ for which $\rho_{AB}$ is
entangled
and thus QPA can be applied:

\beqa
\textstyle{for~ the~ 4-state~ protocol:~ QBER}_{\max}&=&1/4  \label{QBERmax4} \\
\textstyle{for~ the~ 6-state~ protocol:~ QBER}_{\max}&=&1/3  \label{QBERmax6}
\eeqa
Note that these bounds are not only valid for QPA, but more generally for
any quantum
algorithm. Indeed, if the above bounds are violated, then Alice and Bob 
do not share any entanglement, hence their correlation could be
produced by public discussions.

We now discuss the case when Alice, Bob and Eve are left with classical
random variables 
$X$, $Y$ and $Z$, respectively,
and the mutual information Bob-Alice is lower than the Eve-Alice one. Alice
and Bob can then not
achieve a secret key using only error correction and privacy amplification.
Nevertheless, they
can try to use an {\it advantage-distillation} protocol. The idea is that
Alice uses
her random variable $X$ to send over the 
public channel an $N$-bit block encoding a single bit $C$ \cite{effadvdist}:
\beq
X^N\oplus C^N:=[X_1\oplus C,X_2\oplus C,\ldots, X_N\oplus C]
\eeq
Bob then computes $(X^N\oplus C^N)\oplus Y^N$
and (publicly) accepts exactly if this block
is equal to either $[0,0,\ldots,0]$ or $[1,1,\ldots,1]$, corresponding to
$C=0$ and
$C=1$, respectively. In other words,
Alice and Bob make use of a repeat code of length $N$ with only two
codewords $[0,0,\ldots,0]$ and $[1,1,\ldots,1]$. In this way the probability 
that Bob accepts erroneously a bit $C$ goes down like $\D^N$. Eve, on her side,
has to use a majority vote to guess the bit $C$. Hence Bob's information on
$C$ might be
larger than Eve's information even in cases where Bob's information on $X^N$
is lower
than Eve's.
The following theorem defines the possibility for Alice and Bob to achieve 
secret-key agreement using classical algorithms:
\begin{theo}
Secret-key agreement is possible if and only if 
\beq
\frac{\D}{1-\D}<2\sqrt{(1-\delta_0)\delta_0}
\label{necsuf}
\eeq
holds. When applied to the 4- and 6-state quantum cryptography protocols 
\cite{CiracGisin97,FGGNP,Helle6state} this bound
corresponds exactly to conditions (\ref{QBERmax4}) and (\ref{QBERmax6}),
respectively.
\end{theo}
Recall that $\delta_0$ is the probability that Eve guesses correctly the
value of a bit
received undisturbed by Bob.

{\it Proof of Theorem 1:}
First, we note that the condition (\ref{necsuf}) is clearly necessary.
Indeed, for
a disturbance $\D$ larger than this bound the correlations between Alice and
Bob 
correspond to a separable state $\rho_{AB}$ which can be produced without
any quantum
channel, simply by public discussion.

Next, we prove that the condition (\ref{necsuf}) is also sufficient. 
When applying the advantage distillation protocol described above,
Bob's conditional error probability $\beta_N$
when guessing the bit sent by Alice, given that he accepts, is
\beq
\beta_N=\frac{1}{p_{a,N}}\cdot \D^N\leq\left(\frac{\D}{1-\D}\right)^N\ ,
\eeq
where $p_{a,N}=\D^N+(1-\D)^N$ 
is the probability that Bob accepts the received block.
It is obvious that Eve's optimal 
strategy for guessing $C$  is to compute the block
$[(C\oplus X_1)\oplus Z_1,\ldots,(C\oplus X_N)\oplus Z_N]$ and guess $C$ as
$0$ if at least half of the bits in this block are $0$, and as $1$ otherwise.
Given that Bob correctly accepts, Eve's error probability when guessing the 
bit $C$ with the optimal strategy
is lower bounded by $1/2$ times the probability that she decodes
to a block with $N/2$ bits $0$ and the same number of $1$'s. Hence
we get that 
\[
\gamma_N\geq \frac{1}{2}{N \choose N/2}(1-\delta_0)^{N/2}\delta_0^{N/2}
\geq\frac{k}{\sqrt{N}}\cdot\left(2\sqrt{(1-\delta_0)\delta_0}\right)^N
\] 
holds for some constant $k$ and 
for sufficiently large $N$ by using Stirling's formula.
Note that Eve's error probability given that Bob accepts is 
asymptotically equal to her error probability given that 
Bob {\em correctly\/} accepts because Bob accepts erroneously
only with asymptotically vanishing probability, given that he 
accepts.

Although it is not the adversary's ultimate goal to guess the bits 
$C$ sent by Alice, it has been shown 
that the fact that $\beta_N$ decreases exponentially faster than $\gamma_N$
implies that for sufficiently large $N$, Bob has more (Shannon-) information
about the bit $C$ than Eve (see for example~\cite{techrep},\cite{ittrans}). 
Hence Alice and Bob have managed to generate
new random variables with the property that Bob obtains more information about 
Alice's random bit than Eve has. Thus $S(X;Y||Z)>0$ holds for this,
hence also for the original, situation because of the bound (\ref{lbo}).
\pe

In conclusion, we have proven that secret-key agreement using either 4-state
or 6-state quantum
cryptography, assuming general incoherent eavesdropping, 
is possible if and only if the quantum bit error rate is lower than
that produced by the intercept-resend eavesdropping strategy 
($1/4$ and $1/3$ for the 4-state and 6-state protocols, respectively). 
This limit corresponds to complete disentanglement and is valid as well if
Alice and Bob use quantum information processing, like quantum privacy
amplification, as
if they use classical information processing, like advantage distillation
\cite{badQPA}. 
In other words, as long as Alice and Bob share some entanglement, they can
use either
a quantum protocol or a classical protocol to extract a secret-key from this
entanglement.

Quantum Privacy Amplification uses quantum controlled-not gates, the basic
building block
of quantum computers. The latter are usually thought as fundamentally more
efficient than
classical computers. In the case of quantum cryptography our results demonstrate
that the same task can be achieved with both types of computers, but it
remains to be
determined whether one method is more efficient than the other. The case of
coherent
eavesdropping also remains open.

From a practical point of view it is crucial to know the upper error rate
compatible 
with secret-key agreement. It is also important to have good estimates for the 
secret-key rate $S(X;Y||Z)$. Indeed, in practice one has to optimize a
compromise
between high raw bit rates and low error rates \cite{QCexp}.

\small
Stimulating discussions with Artur Ekert, Bruno Huttner, Itoshi Inamori and
Ueli Maurer are
acknowledged. This work was partially supported by the Swiss National
Science Foundation.

\newpage

\section*{Figure Captions}

Fig. 1:  Eve and Bob information versus the QBER, here plotted for
incoherent eavesdropping
on the 4-state protocol. For QBERs below QBER$_0$,
Bob has more information than Eve and secret-key agreement can be achieved using
classical error correction and privacy amplification. These can, in
principle, be implemented
using only 1-way communication. The secret-key rate can be at least as large
as the
information differences. For QBERs above QBER$_0$, Bob has a disadvantages
with respect
to Eve. Nevertheless, Alice and Bob can apply quantum privacy amplification
up to the QBER
corresponding to the intercept-resend eavesdropping strategies, IR$_4$ and
IR$_6$ for the
4-state and 6-state protocols, respectively. 
Alternatively, they can apply a classical protocol called
advantage distillation which is effective precisely up to the same maximal QBER
IR$_4$ and IR$_6$. Both the quantum and the classical protocols require then
2-way communication.
Note that for the eavesdropping strategy optimal from Eve' Shannon point of
view on the
4-state protocol, QBER$_0$ correspond precisely to the noise threshold above
which Bell
inequality can no longer be violated. For the 6-state protocol a similar
relation 
with optimal universal quantum cloning holds \cite{Buzek}.


\begin{thebibliography}{99}

\bibitem{BB84} C. H. Bennett and G. Brassard, in {\it Proceedings of
the IEEE International Conference on Computer, Systems, and Signal
Processing, Bangalore, India\/} (IEEE, New York, 1984), pp. 175--179;
A.E. Ekert, Phys. Rev. Lett. {\bf 67}, 661 (1991);
 see also the Physics World issue of March 1998.

\bibitem{CiracGisin97} I. Cirac and N. Gisin, Phys. Lett. A {\bf229}, 1-7, 1997.

\bibitem{FGGNP} C. Fuchs, N. Gisin, R.B. Griffiths, C.S. Niu and A. Peres, 
           Phys. Rev. A {\bf56}, 1163, 1997.

\bibitem{Helle6state} H. Bechmann-Pasquinucci and N. Gisin, quant-ph 9807041.

\bibitem{Dagmar6state} D. Bruss, Phys. Rev. Lett. {\bf81}, 3018, 1998.

\bibitem{CHSH}J.F.~Clauser, M.A.~Horne, A.~Shimony and R.A. Holt, Phys. Rev.
Lett. {\bf
           23}, 880, 1969.

\bibitem{GisinHuttner97} N. Gisin and B. Huttner, Phys. Lett. A {\bf228},
13-21, 1997.

\bibitem{Buzek} V. Bu\v{z}ek and M. Hillery, Phys. Rev. A {\bf 54}, 1844 (1996);
       N. Gisin and S. Massar, Phys. Rev. Lett. {\bf79}, 2153-2156, 1997.

\bibitem{ka} U.~M.~Maurer, Secret key agreement by public discussion from
common information, {\em IEEE Transactions on Information Theory\/},
Vol.~39, No.~3, pp.~733-742, 1993.

\bibitem{ittrans} 
U.~M.~Maurer and S.~Wolf, Unconditionally secure key agreement 
and the  intrinsic
conditional  information, to appear in  
{\em IEEE Transactions on Information Theory\/}, 1999.

\bibitem{techrep} U.~M.~Maurer and S.~Wolf, Unconditionally secure
secret-key agreement and the intrinsic conditional mutual information,
Tech. Rep. 268, Department of Computer Science, 
ETH Z\"urich, May 1997.

\bibitem{ck} I.~Csisz\'{a}r and J.~K\"orner, Broadcast channels with
confidential messages, {\em IEEE Transactions on Information Theory\/},
Vol.~IT-24, pp.~339-348, 1978.

ref. therein;


\bibitem{QPA} D. Deutsch et al., Phys. Rev. Lett. {\bf77}, 2818, 1996;
       Ch. H. Bennett et al., Phys. Rev. Lett. {\bf76}, 722, 1996.

\bibitem{PeresHorodecki} A. Peres, Phys. Rev. Lett. {\bf77}, 1413, 1996;
       M., R. \& P. Horodecki, Phys. Lett. A {\bf223}, 1, 1996.

\bibitem{effadvdist} Note that there exist much more 
efficient protocols in terms of the amount of extractable secret key.
However, since we only want to prove a qualitative possibility result
here, it is sufficient to look at this simpler protocol.


\bibitem{badQPA} Note that when the disturbance caused by Eve is so low that
she has less (Shannon) information
than Bob on the bits encoded in individual qubits, then neither quantum privacy
amplification nor advantage distillation is needed, classical error
correction and
privacy amplification suffice.
Hence, the terminology {\it quantum privacy amplification} is unfortunate, as it
does not correspond to classical privacy amplification, but include advantage
distillation.

\bibitem{QCexp} H. Zbinden et al., Applied Physics B 67, 743-748, 1998, and
ref. therein;
       G. Ribordy et al., Electron. Lett. 34, 2116-2117, 1998.


\end{thebibliography}
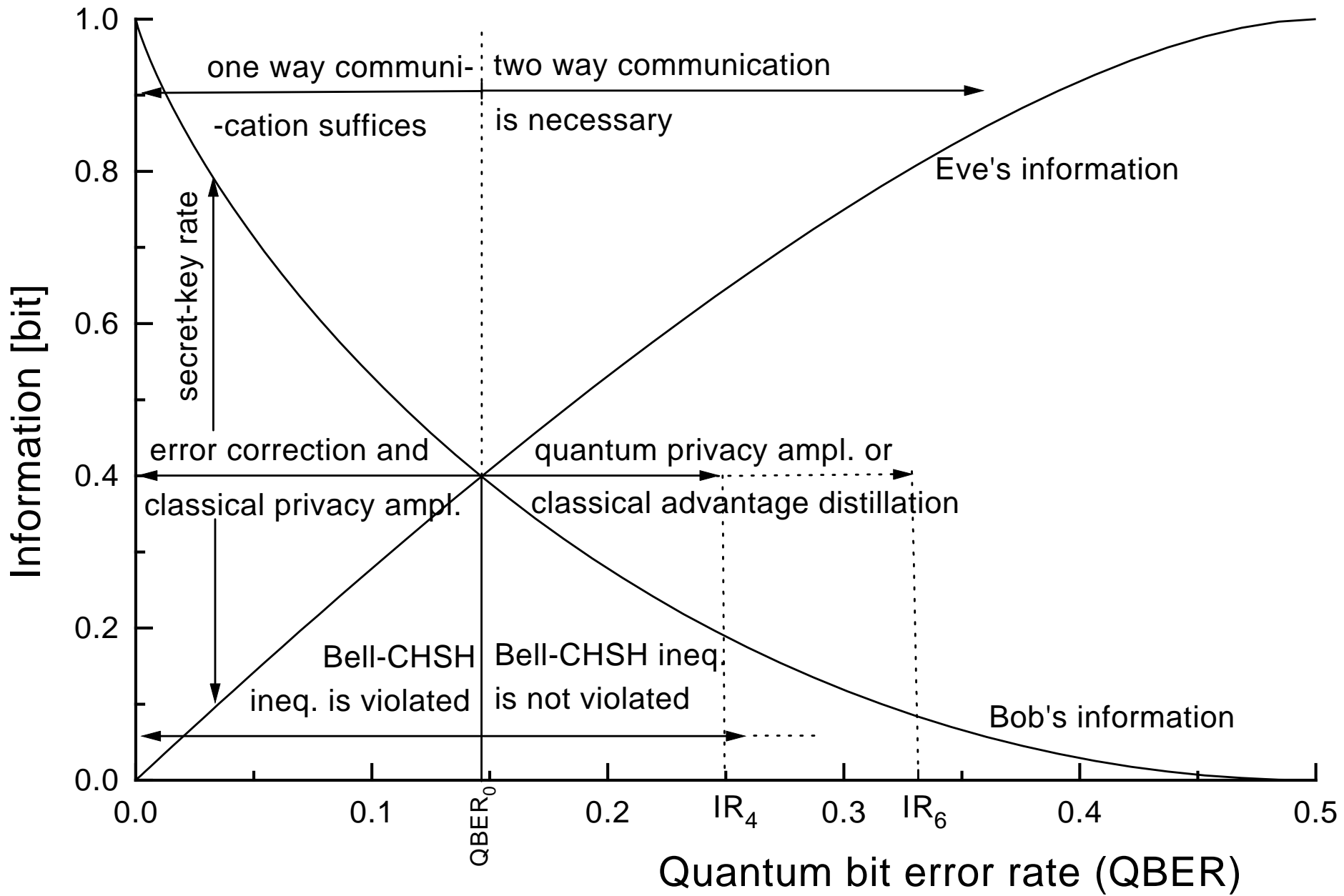
\end{document}